# The Hydraulic Analogues of Basic Circuits: Labs for Online Learning Environments

*J. S. Bobowski,* University of British Columbia, Kelowna, BC, Canada

## Introduction

Due to the COVID-19 outbreak, academic institutions across the globe have been forced to move to online learning environments. It has been particularly challenging for the experimental sciences to develop and deliver authentic lab-based experiences. [1] [2] [3] [4] [5] Some of the strategies that have been adopted for first-year physics labs include: providing a video demonstration and supplying data for students to analyze, using online simulations, [6] [7] [8] [9] having students construct simple apparatuses using commonly-available materials, collecting data with the aid of a smartphone, [10] and, when dealing with smaller class sizes, providing a custom lab kit to each student. [11] Many institutions have chosen to adopt a blend of the various strategies.

In our program, the first semester of first-year physics is mechanics and the second semester is electromagnetism (EM). As much as possible, we wanted students collecting their own data in the lab component of the course. Because the mechanics course is focused on the motion of objects, it was relatively straightforward to design lab exercises that allowed students to collect unique datasets. For example, one of the exercises was to record a block sliding down a ramp and then analyze the captured motion using a video analysis software. Students were able to deduce the coefficients of static and kinetic friction and to show that these quantities were approximately independent of the mass of the block.

It has been much more challenging to design lab exercises for the EM course that meet these same objectives. The material covered is generally more abstract and, perhaps with the exception of a qualitative exploration of static charge, students cannot be expected to use commonly-available materials to investigate the relevant concepts.

In this paper we describe a couple of exercises that we developed for our online first-year EM course. The labs that we describe are focused on basic concepts from circuit analysis, namely Ohm's law and charging a capacitor with a constant voltage source and a series resistor. To give students a visual demonstration of these concepts and, simultaneously, provide a means for students to collect their own unique dataset, hydraulic analogues of these electric circuits were built and a series of videos were recorded. Incidentally, the visual demonstrations may also help to alleviate some of the commonly-held student misconceptions about basic circuits. [12] For this reason, even once in-person labs resume, we expect that the hydraulic analogue demonstrations will continue to be useful.

## Hydraulic Analogues

Hydraulic models for passive circuit elements were described by Bauman in 1980. [13] Although the actual devices were not built, the analogue circuits described, such as the $LC$ resonant tank, offer many useful insights. The hydraulic analogue of a resistor is simply a long, narrow tube through which a fluid flows. The volume flowrate $Q$ is governed by the Hagen-Poiseuille equation:

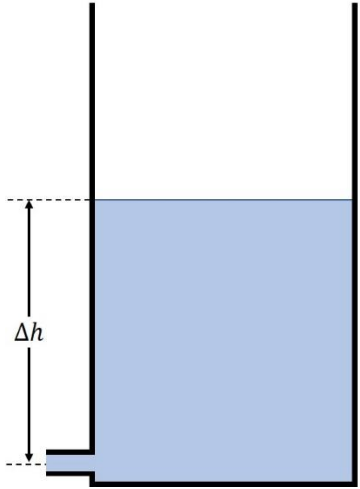

*Figure 1: Schematic diagram of a hydraulic capacitor.*

$$\Delta P = \frac{8\mu L}{\pi r^4} Q, \tag{1}$$

where $\Delta P$ is the pressure difference across the tube, $\mu$ is the dynamic viscosity of the of the fluid, $L$ is the length of the tube, and $r$ is its inner radius. [14] Equation (1) is the analogue of Ohm's law. The pressure difference plays the role of a voltage difference that drives the flow of the "charge", which in this case is the fluid volume. The hydraulic resistance is given by $R_{\mathrm{H}} = 8\mu L/(\pi r^4)$. This expression reveals some of the limitations of the analogue which are useful to keep in mind. First, in the electrical case, the resistance is determined by the resistivity and geometry of the conductor carrying the flow of charge. For fluids, similar to the electrical case, the resistance is determined by the geometry of the tube carrying to fluid. However, the resistance is also determined by the viscosity which is a property of the fluid, not the tube. Furthermore, in the electrical case, $R \propto r^{-2}$ while in the hydraulic case, $R_{\mathrm{H}} \propto r^{-4}$.

In our experiments, we use an open container with an outlet near the bottom as a hydraulic capacitor. It is assumed that the cross-sectional area $A$ of the container is much greater than the cross-sectional area of any of the resistors in the system. Figure 1 shows a partially-charged hydraulic capacitor. By definition, the electrical capacitance is given by the ratio of the stored charge to the voltage difference. Therefore, the hydraulic analogue is $C_{\mathrm{H}} = V/\Delta P$, where $V$ is the volume of the stored fluid. The pressure difference between the open surface of the fluid and the outlet is given by $\Delta P = \rho g \Delta h$, where $\rho$ is the density of the fluid, $\Delta h$ is the height difference, and $g$ is the gravitational acceleration. Combining these results, and using $V = A\Delta h$, results in $C_{\mathrm{H}} = A/(\rho g)$. Notice that the hydraulic capacitance is independent of the height of the container. Although the height does not determine the capacitance of the container, it does set the maximum "charge" or fluid volume that can be stored. In this way, an overflowing hydraulic capacitor is analogous to dielectric breakdown in an electrical capacitor.

If the flowrate from the outlet of a hydraulic capacitor with a large cross-sectional area is low, then the height difference $\Delta h$ remains approximately constant over a relatively long period of time. In this case, the capacitor acts as a nearly constant pressure source, or a hydraulic battery.

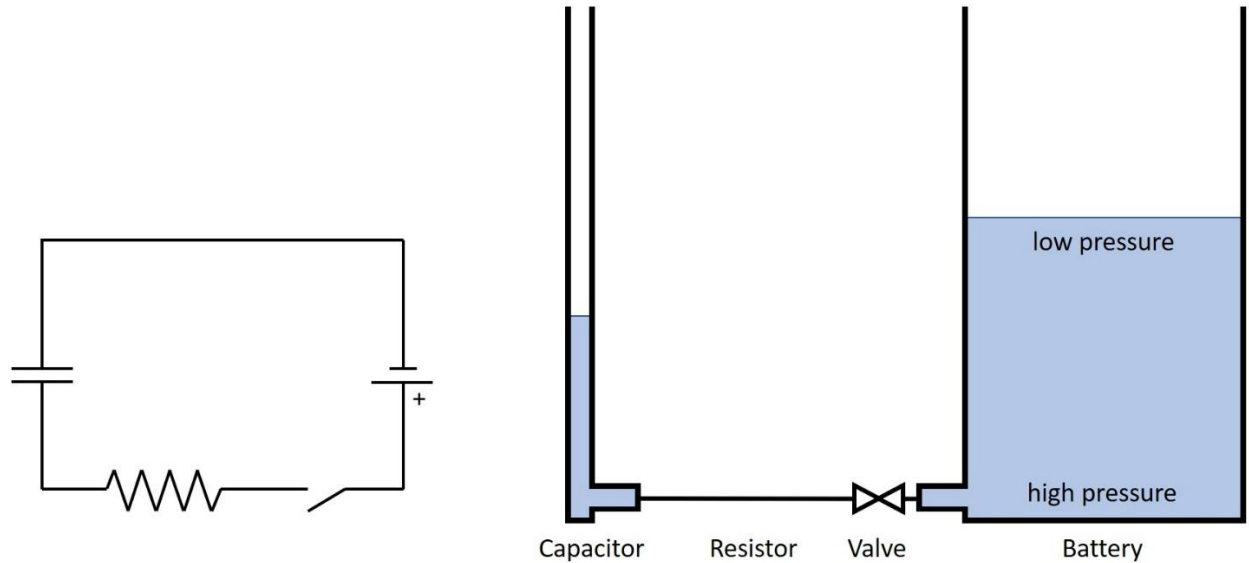

*Figure 2: Series RC circuits. The electric circuit (left) and hydraulic analogue (right). The resistance in the hydraulic circuit is a small-diameter tube.*

As shown in Fig. 2, combing all of these elements allows one to build a hydraulic $RC$ circuit. In the figure, the valve acts as the switch. Note that the high-pressure side of the large container corresponds to the positive terminal of the battery. Also, the hydraulic "circuit" is completed because the surfaces of the fluids in the battery and capacitor are at the same atmospheric pressure (i.e. the same "voltage"). The time constant associated with the hydraulic circuit is given by $\tau_{\mathrm{H}} = R_{\mathrm{H}} C_{\mathrm{H}}$ or, equivalently:

$$\tau_{\mathrm{H}} = \frac{8\mu L}{\pi r^4} \frac{A}{\rho g}. \tag{2}$$

## Experiments

The experimental realizations of the hydraulic circuits are shown in Fig. 3. A large glass container with an outlet at the bottom served as the constant pressure source (battery). The glass container was about 35 cm tall and holds approximately 10 L of liquid. The container was filled with tap water that had two squirts of hand soap added to act as a surfactant. The addition of soap suppresses surface tension effects and allows the water solution to penetrate into the narrow capillaries used as resistors. It also helps to minimize the meniscus that forms at the surface of the liquid in the glass tube that was used as the capacitor. We added food coloring to the solution so that the height of the fluid in the glass tube was easily tracked. A ruler was placed next to the capacitor so that the liquid's height could be measured with a resolution of about 0.5 mm. The capacitor tube was about 70 cm long and had an inner diameter of 5.85 mm. For the resistors, we used various polyethylene and Teflon tubes. The tube diameters were measured by taking a photograph of the cross-sections using a calibrated microscope and then analyzing the images in the open-source image-processing software ImageJ. The inner and outer radii of the resistor tubes used are given in Table 1. All resistors had a length of 94.5 cm.

To connect the resistors to the battery, a small hole was first drilled though a rubber stopper. A thin coating of vacuum grease was added to one end of the capillary tubing and then it was inserted into the hole until it passed completely through the rubber stopper. The stopper was then pressed into the outlet of the glass container. For the Ohm's law experiment, open end of the resistor tube was held at the same height as the battery outlet and allowed to drip into an open beaker placed on a digital scale. The scale measures the mass flowrate and, therefore, acts as an ammeter. Dividing the mass flowrate by the fluid's density yields the volume flowrate $Q$.

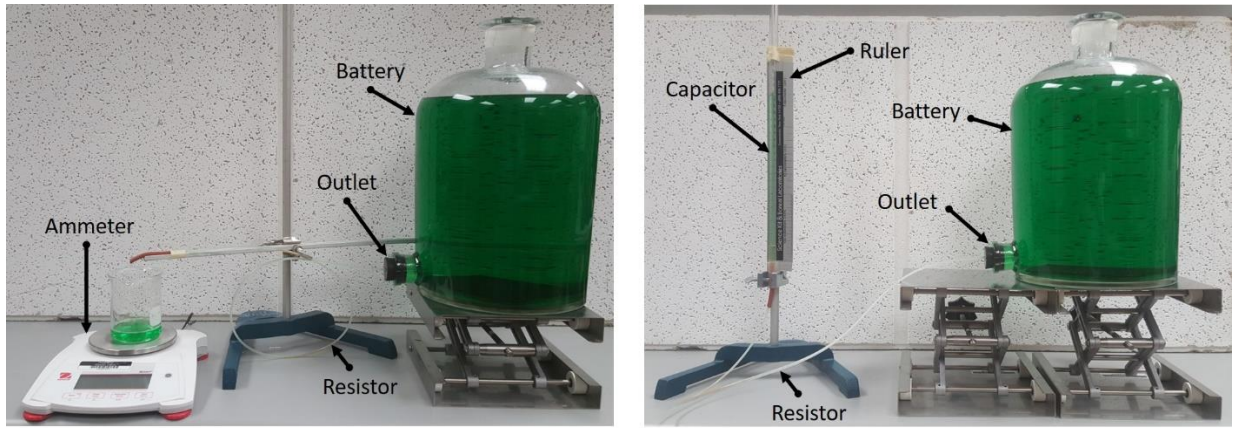

*Figure 3: Experimental setups for the hydraulic analogues of Ohm's law (left) and charging a capacitor (right).*

| Resistor Number | Inner radius (mm) | Outer radius (mm) | $r_{\text{eff}}$ (mm) |
|---|---|---|---|
| 1 | 0.99 | 1.53 | 0.99 |
| 2 | 0.64 | 0.97 | 0.64 |
| 3 | 0.21 | 0.46 | 0.28 |

*Table 1: Inner, outer, and effective radii of the hydraulic resistors.*

For the charging experiment, the open end of the resistor tube was connected to the bottom of the capacitor using a series of tight-fitting rubber tubes of increasing sizes. The bottom end of the capacitor was set to be level with the battery outlet. Rather than using a valve to start the flow of liquid as shown schematically in Fig. 2, we applied a few psi of air pressure at the top of capacitor tube to completely discharge the capacitor. The charging process was then initiated by removing the excess pressure.

## Results

We now present the experimental results.

### Ohm's law

The difference in height between the fluid surface and the outlet of the battery was measured to be $\Delta h = 23.1 \pm 0.4$ cm. Assuming a density of $\rho = 997 \text{ kg/m}^3$, the constant pressure difference supplied by the battery was $\Delta P = 2.26 \pm 0.04$ kPa. The volume-versus-time measurements are shown in Fig. 4. The videos provided to the students, and used to obtain the data in Fig. 4, were uploaded to YouTube and are available for anyone to use. [15] The lines in the figure are linear fits and the slopes determine the volume flowrate $Q$. Table 2 summarizes the measured and calculated values of $Q$ and $R_{\text{H}} = \Delta P / Q$, respectively.

### Annular and parallel flowrates

To adjust the hydraulic resistance in our experiments, we inserted the smaller-diameter capillaries into the larger-diameter capillaries. This strategy was convenient because we did not have to make watertight seals for each individual resistor. However, it adds the complication that the net volume flowrate is the sum of the flowrate through the inner diameter of the smallest tube and the flowrate through the annual region between concentric tubes. As shown in Table 1, the outer diameter of tube #2 is closely matched to the inner diameter of tube #1. Therefore, annular flow only affects the measurements when tube #3 is inserted into tube #2.

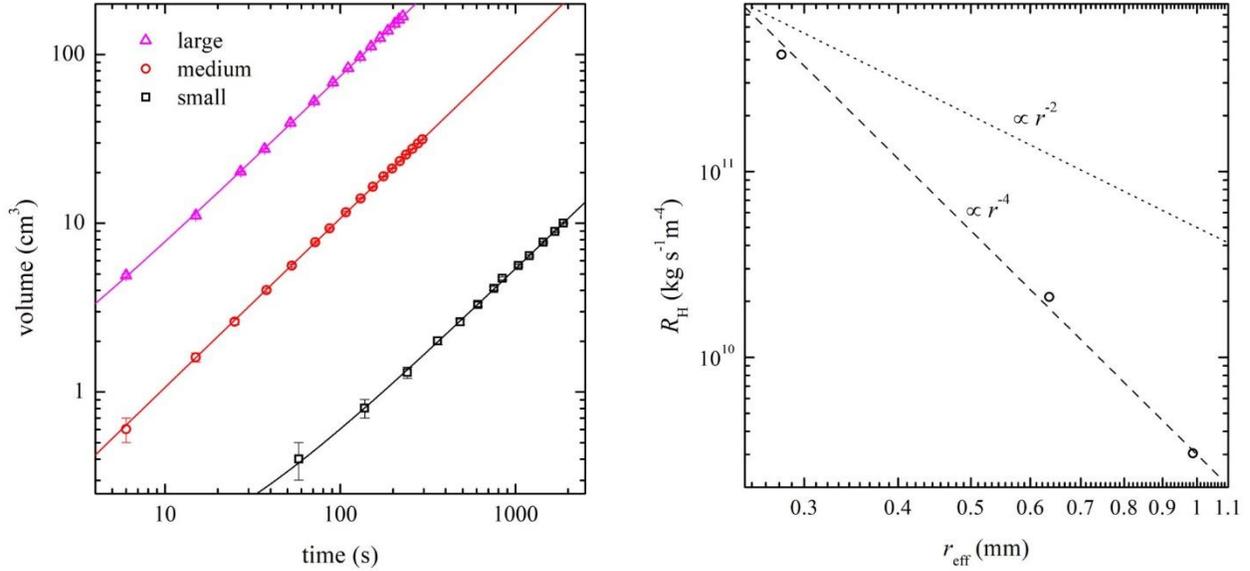

*Figure 4: Left: Volume versus time data used to determine the volume flowrate. Right: Calculated hydraulic resistance as a function of the effective resistor radius. The error bars are smaller than the datapoints.*

| Tube Number | $Q$ (cm$^3$/s) | $\Delta P$ (kPa) | $R_H$ ($10^9$ kg s$^{-1}$m$^{-4}$) | $\tau_H$ (s) | $C_H$ ($10^{-9}$ m$^4$s$^2$/kg) |
|---|---|---|---|---|---|
| 1 | 0.741 | 2.26±0.04 | 3.05±0.05 | 7.29±0.04 | 2.39±0.04 |
| 2 | 0.107 | 2.26±0.04 | 21.1±0.4 | 52.8±0.2 | 2.50±0.05 |
| 3 | 0.00531 | 2.26±0.04 | 426±8 | 962±3 | 2.26±0.04 |

*Table 2: Summary of the hydraulic resistance and capacitance determinations. The relative error in $Q$ is negligible compared to the relative errors in the other tabulated quantities.*

The equivalent of the Hagen-Poiseuille equation for annular flow is:

$$\Delta P = \frac{8\mu L}{\pi} \left[ (r_2^4 - r_1^4) - \frac{(r_2^2 - r_1^2)^2}{\ln(r_2/r_1)} \right]^{-1} Q, \tag{3}$$

where $r_2$ is the outer radius of the annular region and $r_1$ is its inner radius. [16] If there are parallel flows due to an annular region and a central small tube of inner radius $r_0$, then the net flowrate is given by:

$$Q_{net} = \frac{\pi \Delta P}{8\mu L} \left[ r_0^4 + (r_2^4 - r_1^4) - \frac{(r_2^2 - r_1^2)^2}{\ln(r_2/r_1)} \right] \equiv \frac{\pi \Delta P}{8\mu L} r_{eff}^4, \tag{4}$$

where $r_{eff}^4$ is defined to be everything contained within the square brackets in Eq. (4). The value of $r_{eff}$ for the hydraulic resistors used are given in Table 1. Annular flow affects only resistor #3 and it results in an effective radius that is about 30% greater than the inner radius of the smallest tube. However, because $Q \propto r^4$, the resulting flowrate approximately triples.

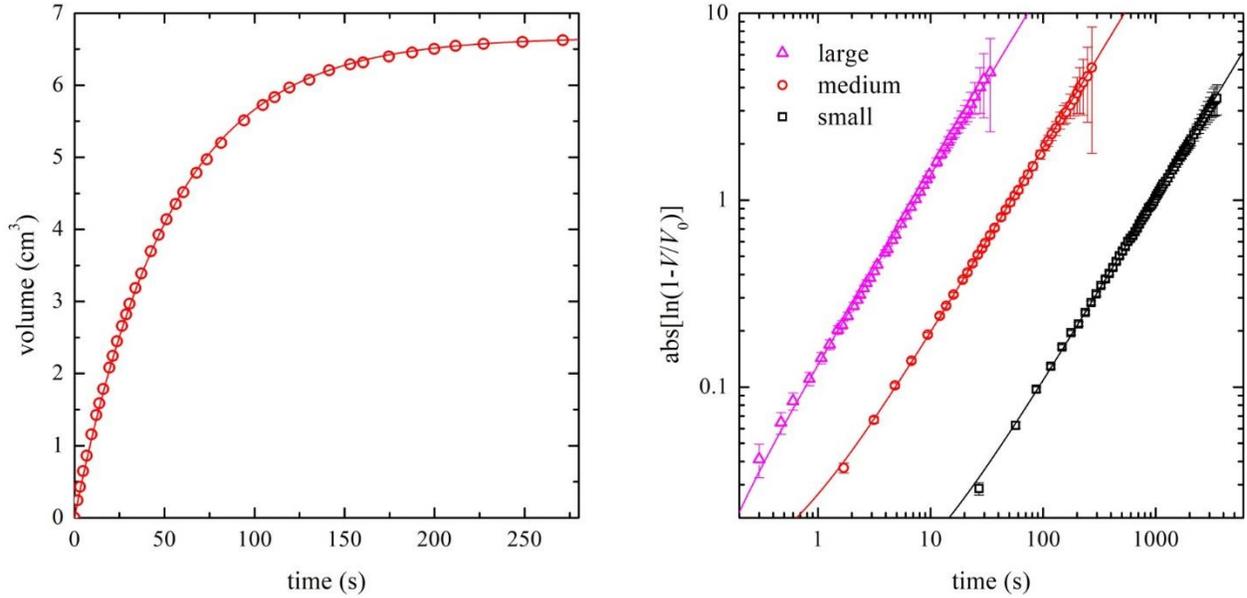

*Figure 5: Charging a hydraulic capacitor. Left: Stored fluid volume as a function of time for resistor #2. The error bars are smaller than the datapoints. The line is the theoretical model using the time constant obtained from a fit to the linearized data. Right: Linearized charging data for the three different resistor tubes. The lines are linear fits to the data.*

Figure 4 shows a plot of the hydraulic resistance versus $r_{\text{eff}}$ on a log-log scale. As expected, the three data points follow a $r^{-4}$ trend. The graph also shows a $r^{-2}$ trend line that would be expected for electrical resistors. We also note that it is in principal possible to use the $R_{\text{H}}$ measurements and the geometry of the tubes to determine the dynamic viscosity $\mu$ of the fluid. [17] However, this measurement is challenging because, for example, a 10% error in tube radius will lead to a 40% error in the determination of $\mu$.

**Charging a capacitor**

An example volume versus time dataset when using resistor #2 is shown in Fig. 5. The videos used to acquire the data for all three resistors are also available on YouTube. [18] The line in the figure was made using the theoretical model expected for a charging capacitor:

$$V = V_0\left(1 - e^{-t/\tau_{\text{H}}}\right), \tag{5}$$

where $V_0$ is the final equilibrium volume and $\tau_{\text{H}} = R_{\text{H}}C_{\text{H}}$ is the charging time constant. Equation (5) can be linearized such that:

$$\ln\left(1 - \frac{V}{V_0}\right) = -\frac{t}{\tau_{\text{H}}}, \tag{6}$$

and $\tau_{\text{H}}$ can be determined for the slope of a linear fit. Figure 5 shows $\text{abs}[\ln(1 - V/V_0)]$ plotted as a function of time for all three resistors on a log-log scale. The lines are linear fits to the data and the extracted time constants are given in Table 2. The table also gives the experimentally-determined hydraulic capacitance calculated from $C_{\text{H}} = \tau_{\text{H}}/R_{\text{H}}$. The experimental results are lower than the expected value of $A/(\rho g) = 2.75 \times 10^{-9} \text{ m}^4\text{s}^2/\text{kg}$ by, on average, about 14%.

One possible reason for this discrepancy could be an over estimate of the pressure difference $\Delta P$ used to calculate $R_\text{H}$ from $Q$. As can be seen in Fig. 3, the glass container used as a battery does not have completely vertical walls. The jar is narrower at the top than it is at the bottom. Therefore, the pressure difference is less than $\rho g \Delta h$. Furthermore, it is difficult to accurately measure the difference in height between the surface of the fluid and the position of the resistor tube that penetrates the battery outlet. Finally, in both the Ohm's law and capacitor charging experiments, it is difficult to ensure that the two ends of the resistor tubes are at precisely the same height.

## Summary

We have developed hydraulic circuits to experimentally investigate Ohm's law and the charging time of a capacitor. Although hydraulic circuits and circuit components have previously been reported in the physics education literature, [19] [20] [21] to the best of our knowledge, the work that we have described is unique. Our work was motivated by the need to develop lab projects for large first-year EM courses that are delivered entirely online. The projects provide a visual depiction of common phenomena from electric circuits. The recorded videos allow each student to collect unique datasets that they then analyze. The data acquired is clean and follows the theoretical models for Ohm's law and a charging capacitor very well. We also note that industrious students ought to be able to construct equivalent apparatuses using only items that are relatively easy to acquire.

The videos of the experiments are freely available for anyone to use. [15] [18] In the online labs that we deploy, the measurements described in this paper will be supplemented with electric circuit simulations from PhET and/or oPhysics. [22] [23]